# A scatter correction method for Multi-MeV Flash Radiography

Jia Qinggang*[1], Mao Peng-Cheng[1] Yang-Bo[1] Shi Dun-Fu[1] Zhang Ling-Yu[1] Deng-Li[1] Xu Hai-Bo[1]

***Abstract*—Multi-MeV flash radiography is often used as the primary diagnostic technique for high energy and density (HED) physics experiments. Matter under extreme conditions such as metal with high density produced by implosion is the main study object of HED experiments. Primary X-ray which is attenuated by the object offers density information of the object. For a thick metal object with area density as high as 150 g/cm$^2$, the rest part of primary X-ray which passes through the object may drowned in scattered X-ray fog. It seriously limits accuracy of density quantification. In general, High-Z anti-scatter grid for scatter rejection is used in flash radiography, and a thicker anti-scatter grid works better than a thinner one for Multi-MeV X-ray. However, one essential condition using this anti-scatter grid should be guaranteed that the position of X-ray source is stable and located at the same point which the multichannel grid focus. Otherwise, the anti-scattering grid functions as a shield instead of a collimator for both primary and scatter X-ray, and the costly HED experiment takes risk of getting a useless X-ray image. In this research, an online scatter estimation method is newly designed which can be easily arranged by putting an additional slit collimator downstream of the general X-ray radiography layout. The basic ideal of this method is that the proportion of scatter and primary x-ray will be changed a lot when x-ray passes through a slit like collimation, then scatter component is solvable with known the collimation performance of the silt on scatter and primary x-ray. Monte Carlo simulation shows that, with this method, evaluation error of average scatter is less than 2% when object area density is as high as 200 g/cm$^2$. In addition to the average scatter, an accelerated Monte Carlo method is developed to obtain scatter distribution in iterative reconstruction. By employing the Genetic algorithm as an optimizer, reconstruction can be done by searching a density which projection with scatter best matches experimental result. This reconstruction method requires neither a priori-knowledge such as mass restriction nor regularization. Simulations show that for France Test Object (FTO), the error of reconstructed density is less than 2%, and uncertainty basically covers the real density.**

## I. INTRODUCTION

The HED experiment mainly focus on physical phenomenon of the matter at extreme conditions. The object matter of HED often has very high density (with area density over 100 g/cm$^2$). To measure the object at such high density, Mega-volt X-ray nondestructive testing method is generally employed. Similar to medical X-ray radiography, the density information of object can be obtained according to the attenuation of primary X-ray. When a narrow X-ray beam traverses an object, the beam intensity is reduced caused by absorption or by Compton scattering of X-ray photon from beam. In the high-density object, most of MeV X-ray photons that are scattered will scatter in a forward direction and may captured by scintillation imager. This scatter X-ray during an exposure results in fog on the radiograph. The scattering induced fog is neither a constant nor a random noise on the captured image. It has relation with radiography layout [1,2] and the object. For a thick object, such as FTO, the area density is over 200g/cm$^2$. Monte Carlo simulation of 4MeV X-ray flash radiography about FTO object, in Fig.1.a, shows that the ratio of primary to scattered x-ray on image may be less than 0.1. Therefore, without scatter correction, direct reconstruction from captured image can hardly offer real density, as the result is shown in Fig.1.b.

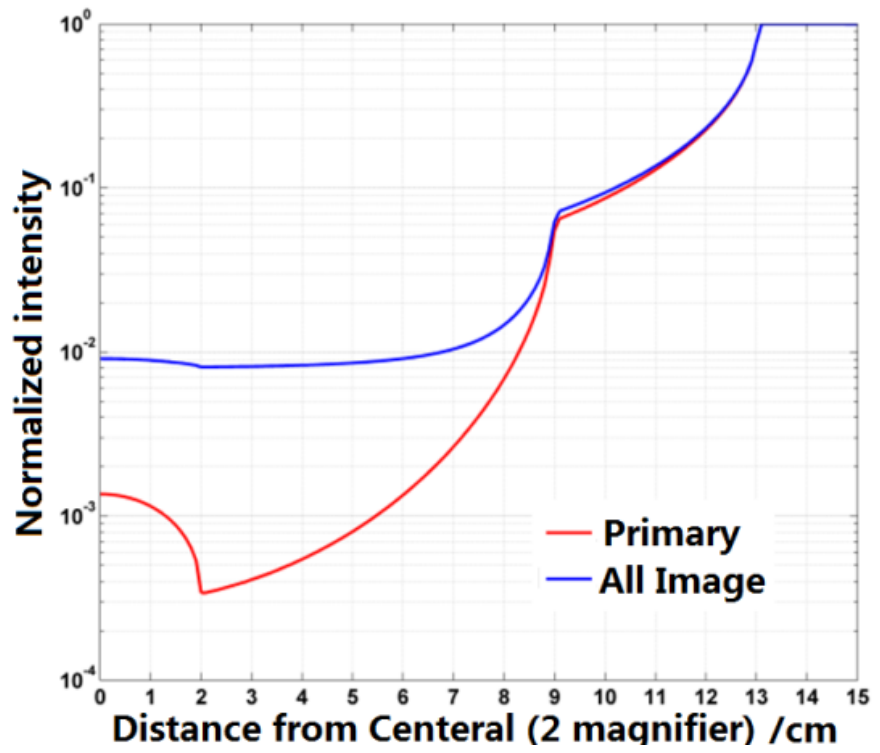

a） Expected visible light distribution and primary component of FTO

This work was supported in part by the National Natural Science Foundation of China ()

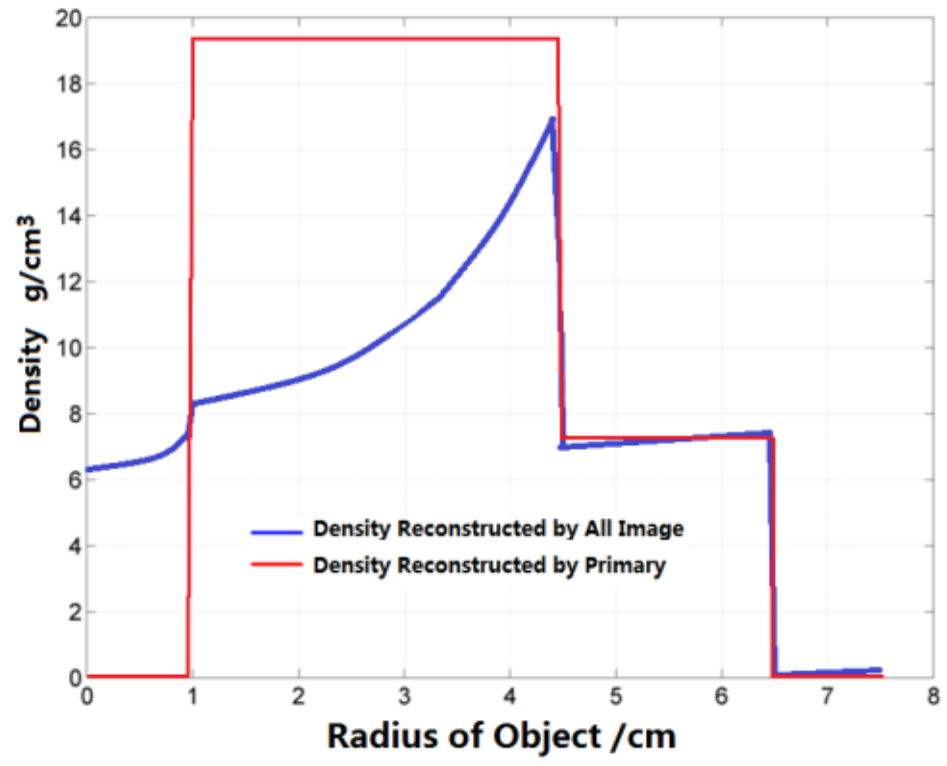

b） Reconstructed densities by visible light

Fig.1. Scatter and primary X-ray distribution of FTO and density reconstruction

For medical X-ray imaging purpose, many papers were published related to scatter correction. The studied suppression methods[3,4] are generally in the terms of hardware techniques for scatter rejection and software algorithms of scatter correction .

In the research field of HED flash radiography, collimators near source restricting Field-of-View (FOV) and at detector side rejecting scatter are all employed. In radiography experiment carried out at Dual Axis Radiographic Hydro Test (DARHT[5]) facility utilizes a thick anti-scatter grid[6] made of high-Z tungsten to collimate MeV x-ray. All sub-tubes of this grid should focus the same point where x-ray source located. This grid is very expensive, and hard to be manufactured. In addition to this, the grid thickness and sub-tube diameter should be carefully designed. A better collimation performance prefers a grid with higher ratio of thickness over sub-tube diameter. But for such grid, the focus area in which x-ray source can see the detector through every sub-tube will be limited in a very small region. The accelerator-driven MeV x-ray source spot may easily shifts out of grid focus, not only scattered x-ray, but also uncollided primary x-ray will be shielded by the grid, and the experiment scarcely obtains a useful image.

In terms of software, algorithm related to scatter correction, in general, has two components, scatter estimation and scatter compensation in reconstruction. In the research field of HED flash radiography, Cao[7] proposed a scatter iterative correction method. In this method, estimations of primary and scattered x-ray is calculated by characteristic ray and pre-obtained Monte Carlo simulations respectively. Liu[2] has studied a scatter correction approach with FTO radiography experiment. This approach also comprises scatter estimation and scatter compensation procedure in reconstruction. The scatter estimation of FTO by Monte Carlo was validated with experimental result. Mass restriction on reconstructed object was employed to evaluate overall scale constant C which offers a link between simulated scatter distribution and experimental one. However, mass restriction may not works well for layers closed to the central of the object, because the volumes of these layers are small and changes in density can barely affects total mass. No further experimental studies are found which shows the performance of mass restriction on determining scale C with various thick objects.

In this paper, a new scatter compensation method is purposed. It comprises two parts: a scatter estimation and a scatter correction. For scatter estimation, this method combines an novel measurement design about online average scatter and an accelerated Monte Carlo simulation for scatter distribution. Section 2 will show scatter evaluation part. In Section 3, a reconstruction procedure based on Bayes framework is developed for scatter correction. Neither a priori-knowledge constraint such as mass restriction nor regularization is introduced in this reconstruction procedure. The reconstructed result and corresponding uncertainty is presented. In Section 4, we give the results and discussions. Conclusion are given in the last section.

## II. Scatter Evaluation Method

In x-ray radiography, captured visible image( $I_{Im}$ ) is contributed by primary ($I_{Uc}$) and scattered x-ray ($I_{Sc}$). The intensity of primary x-ray can be used to calculate mass thickness of object in form of Beer–Lambert Law

$$\frac{I_{Uc}}{I_0} = \exp(-\int_0^L \Sigma_k \rho_k dL) \quad (1)$$

where $I_0$ is primary x-ray intensity before attenuation is primary x-ray intensity, $\Sigma_k$ is the average line absorption coefficients of k-district in object along the x-ray beam, $\rho_k$ is density of k-district, the integration in exponent function gives optical length of x-ray. The scatter part $I_{Sc}$ should be removed ahead of spatial density reconstruction. Many analytic or mathematical models[8-10], such as Beam-scatter-kernels ,are well used to study scatter in medical imaging. In recent years, Deep Neural Network [11,12] is trained based on Monte Carlo simulation results for real-time scatter estimation. Although, mathematical physics model, Monte Carlo and machine learning are well employed for scatter, direct online measurement is superior for experiment.

Here, a new scatter measurement approach is proposed and the layout can be found in Fig.2. The basic ideal of this approach is that the proportion of scatter and primary x-ray will be changed a lot when x-ray pass through a slit collimation, then scatter component can be calculated.

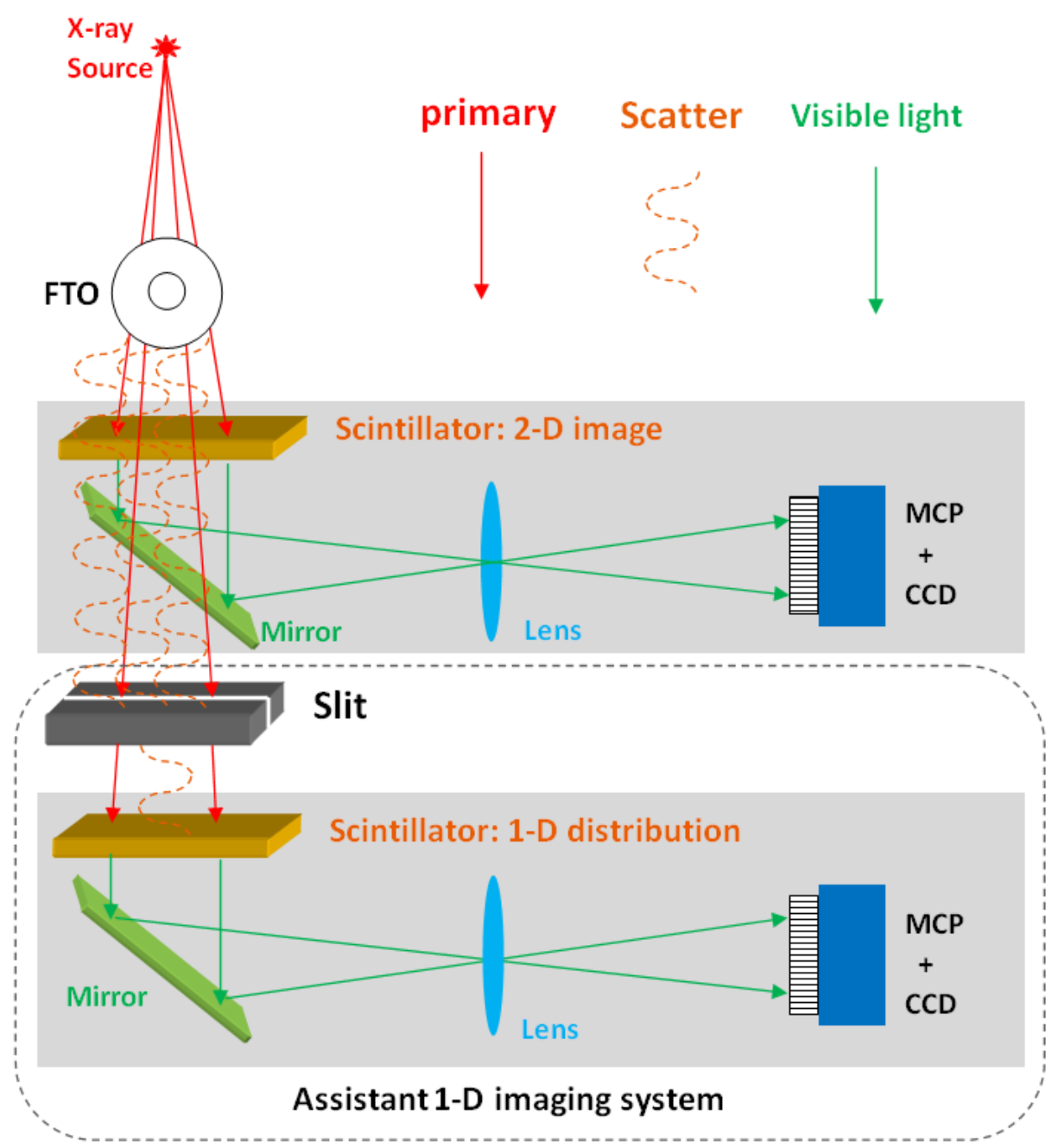

Fig.2. layout of a general x-ray imaging system coupling with newly proposed assistant 1-D imaging system

As shown in Fig.2, a general x-ray imaging system has two main parts. The first is scintillator which transfers x-ray to visible light. The second is optical imager which is composed of a visible light reflector, a lens, and a optical imager(MCP+CCD). The proposed assistant 1-D imaging system is similar to the general imaging system, difference can be found that a slit (or similar shape) collimation is arranged downstream the x-ray beam. This slit has two functions, one is blocking the scattered x-ray, another is offering a path for primary x-ray without attenuation. Therefore, the proportion of scatter on the assistant image is largely reduced compared to that on the general image. As a comparison, the intensity of primary x-ray is slightly changed when it penetrates scintillator and mirror. Therefore, the scatter proportion can be calculated from the general and assistant image. Here two FTO shape object with different density are taken to show the calculation procedure based on this proposed method. The two objects are named, according to the maximum density, as FTO-U25 and FTO-U35,respectively. These density distributions with radius can be found in Fig.3.a.

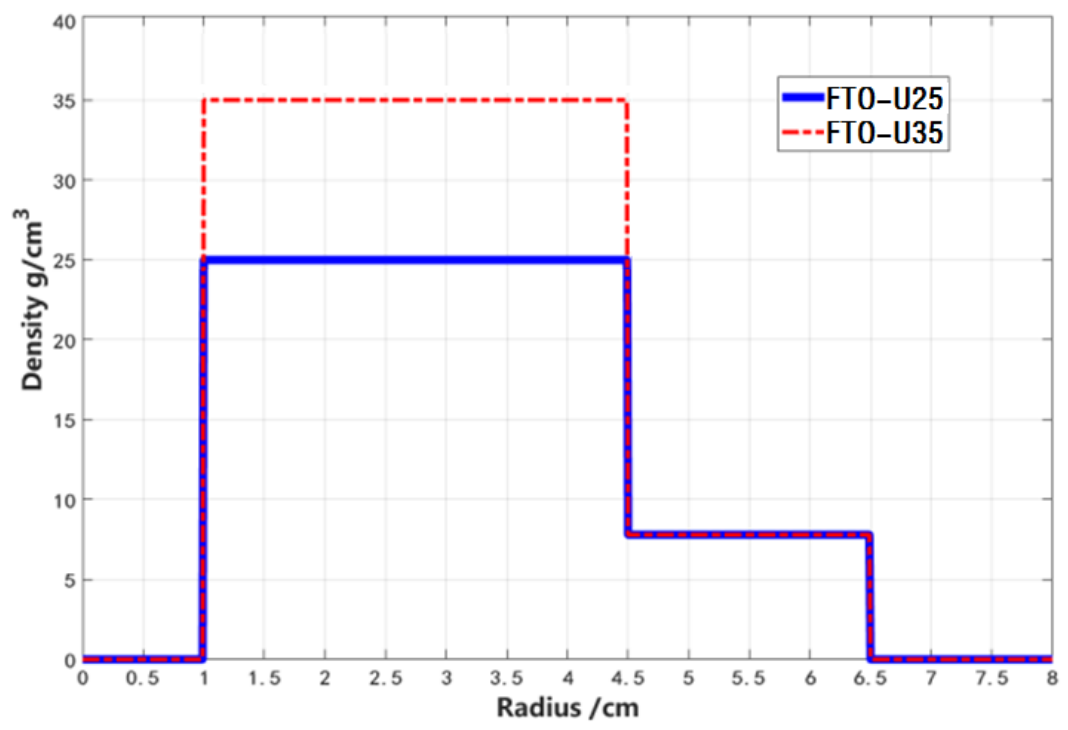

a) Density varies with radius

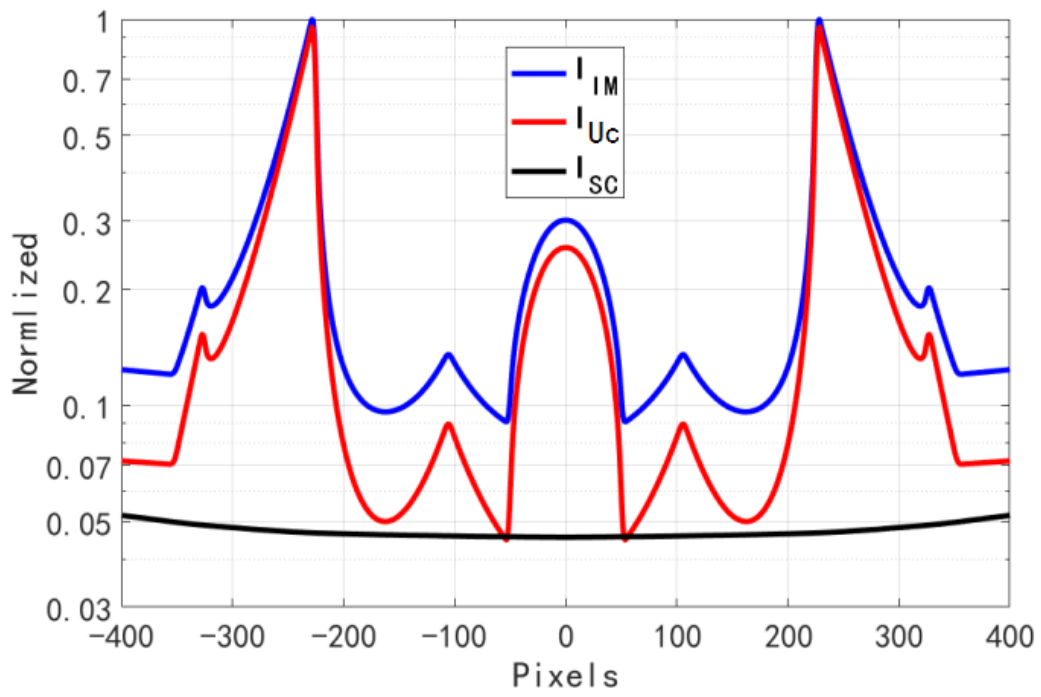

b) Expected visible light distribution $I_{Im}$ and its primary $I_{Uc}$ and scatter $I_{Sc}$ components

Fig.3 Density distribution of FTO shape objects and simulated results of FTO-U25

Monte-Carlo simulation was carried out to calculate the scatter and primary x-ray intensity captured by general imaging system. Results of FTO-U25 is shown in Fig.3.b. At region of pixel with number 230(or -230) , primary intensity is over one order higher than that of scattered x-ray. While, at the place near pixel 50, the corresponding area density projected by FTO is almost the highest, and scatter x-ray is same to that of primary.

Theoretically, only with visible light image $I_{Im}$ which is captured by general imaging system, it is impossible to get $I_{Uc}$ and $I_{Sc}$ by solving equation 2:

$$I_{Im} = I_{Uc} + I_{Sc} \qquad (2)$$

If another independent equation corresponding to $I_{Uc}$ and $I_{Sc}$ can be obtained, both of them are resolvable. Here, a new proposed assistant 1-D imaging system may provide such necessary equation as shown in equation 3:

$$I_{AIm} = I_{AUc} + I_{ASc} \qquad (3)$$

where, $I_{AIm}$ is visible light image captured by assistant 1-D imaging system, $I_{AUc}$ and $I_{ASc}$ is primary and scatter component of $I_{AIm}$, respectively.

$I_{AUc}$ is in proportion to $I_{Uc}$. The ratio of $I_{AUc}$ to $I_{Uc}$ named K(<1) is determined both by space efficiency related to imaging layout and by attenuation of the scintillator in the general imaging system. Monte Carlo simulation shows that K is a constant in the FOV. Since the scattered x-ray is reduced by the slit, $I_{ASc}$ is smaller than $I_{Sc}$. Ratio of $I_{ASc}$ over $I_{Sc}$ named R represents suppression performance of the silt collimation on scatter. Value of R is position dependent on the image. With the same slit, distribution of R scarcely varies with different objects. Here, for brief, R is assumed uniformity which equals $R_{ave}$ in the FOV. The scatter x-ray $I_{Sc}$ can be can be calculated by equation (4):

$$I_{Sc} = \frac{K * I_{Im} - I_{AIm}}{K - R_{ave}} \qquad (4)$$

where $I_{Im}$ and $I_{AIm}$ are images measured online. $R_{ave}$ and K can be calibrated ahead with an known object. Although, these two parameters can be simulated by Monte Carlo, measurement is the first choice.

The object FTO-U25 is used for calibrating Rave and K. FTO-U35 is taken to show the scatter evaluation performance. A tungsten slit collimator is utilized which has 10cm length in thickness and 5mm in opening width. Corresponding simulated $I_{Im}$ and $I_{AIm}$ for FTO-U25 are shown in Fig.4. Formula for K is shown as follows:

$$K = \frac{I_{AIm}}{I_{Im}} + \frac{K^0 - R^0}{1 + \frac{I_{Uc}}{I_{Sc}}} \qquad (5)$$

where, $K^0$ and $R^0$ are initial guess. To calibrate K simply, only the first part of right side in equation 5 is used. It is because the expected $I_{Uc}/I_{Sc}$ at pixel 230 is over 20, according to the simulation results shown in Fig.3.b. The errors induced by truncation is at several percents level. The directly calculated K by $I_{AIm}/I_{Im}$ is 0.256.

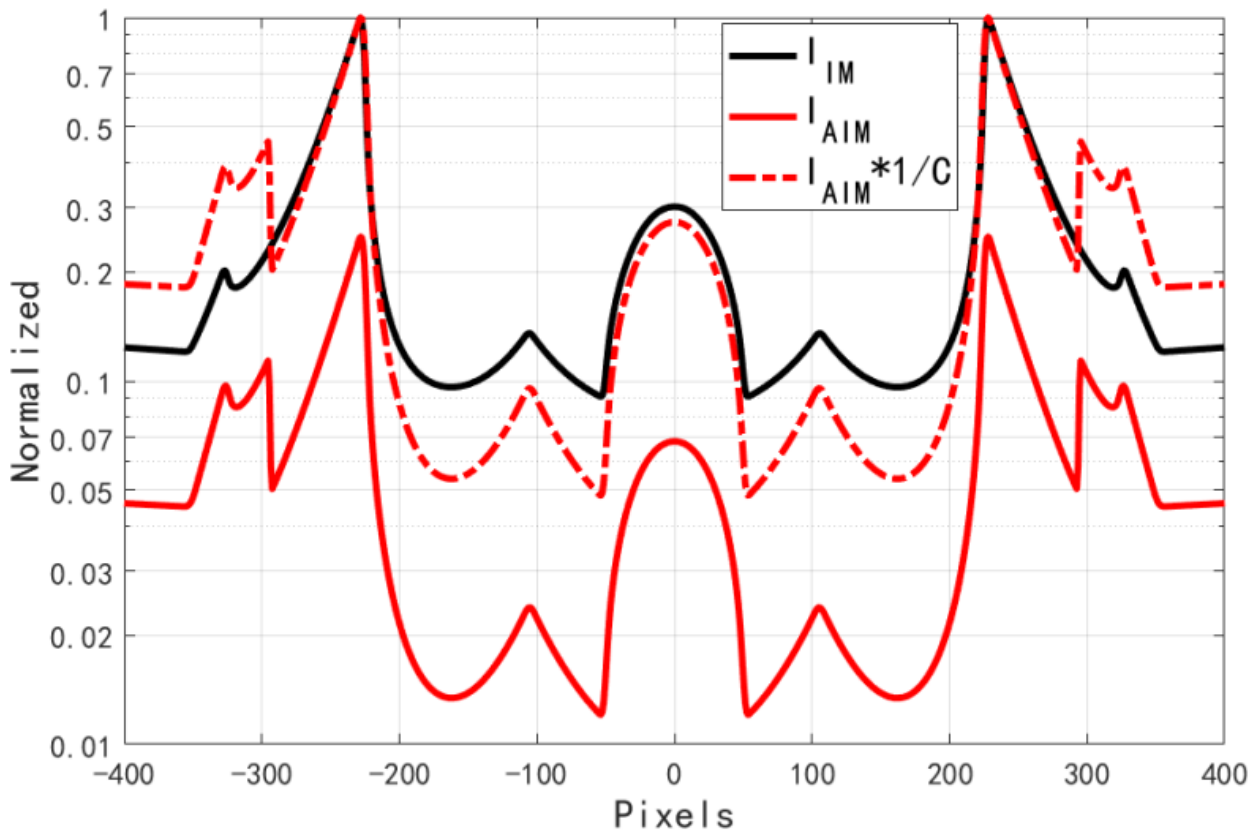

Fig.4 Expected images of FTO-U25 obtained by general and assistant system

In terms of $R_{ave}$ calibration, equation 4 is employed inversely. Having known the density of object such as FTO-U25, $I_{Sc}$ can be well evaluated. It is because that when the reconstructed density matches the object, the input scatter $I_{Sc}$ for correction is approximating to the actual value. Here, the value of $I_{Sc}$ selects Monte Carlo result which value is 0.0455 as shown in Fig.3.b. Then $R_{ave}$ is calculated with value of 0.122. Iteration update with parameter $K^0$ and $R^0$ about the truncated part in equation 5 can be applied for better accuracy. After several iterations, sufficient convergence is achieved, and the improved K and $R_{ave}$ equals to 0.262 and 0.115, respectively.

FTO-U35 is taken to show the scatter evaluation performance. Simulated $I_{Im}$, $I_{Sc}$ and $I_{AIm}$ are given in Fig.5. With K and $R_{ave}$ , the estimated $I_{Sc}$ by equation 4 is about 0.0434 as shown in Fig.5. The Monte Carlo simulated $I_{Sc}$ as the real standard one has a value of 0.0426 at pixel 0 and 0.0433 at pixel 200. The error of scatter estimation in central of FOV is less than 2%.

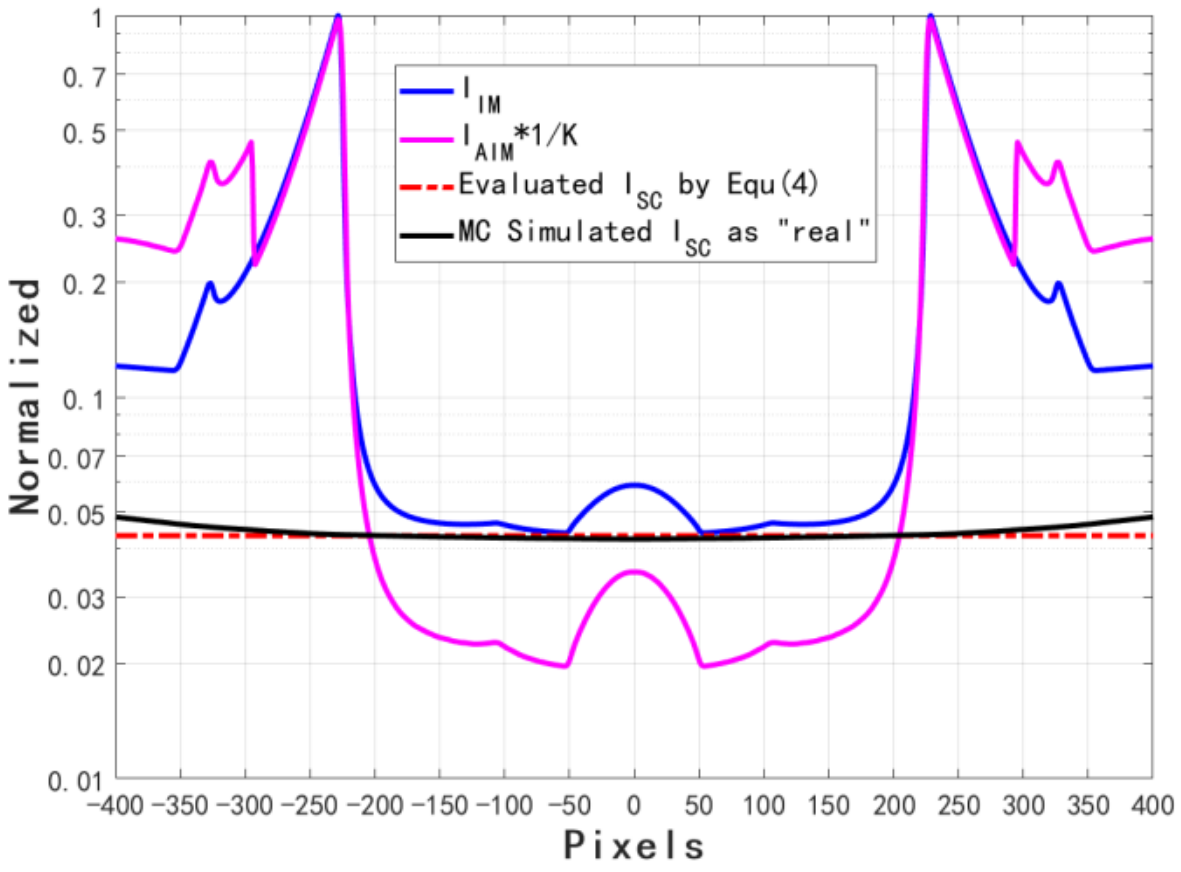

Fig.5 Expected general, assistant images and scatter estimation of FTO-U35

To note that, as shown in equation 4, error of K will transfers to the calculated $I_{Sc}$, especially when value of $I_{Im}$ is high, such for the region near pixel 230 shown in Fig.4. Therefore, equation 4 is suitable for the position at which $I_{Im}$ reaches its minimum.

In the cases given above, the expected scatter distribution is rough flat on the FOV. This condition can be achieved by enlarge the air gap between detector and object[13]. For the cases that the scatter has a sharp distribution, a suitable method is developed which combines this proposed evaluation procedure and Monte Carlo simulation. In this method, Monte Carlo is employed to calculate scatter distribution based on density of object in reconstruction iteration, and the proposed evaluation procedure mentioned above is used to provide scale constant C which plays role of a link between simulated distribution and experimental one. This method will be introduced in next section.

## III. Scatter Correction and Reconstruction Method

When expected scatter is nearly constant, the structure of scatter correction schemes within reconstruction procedure can be found in Fig. 6. All the details about scatter estimation based on general system and newly proposed 1D assistant imaging system are given above.

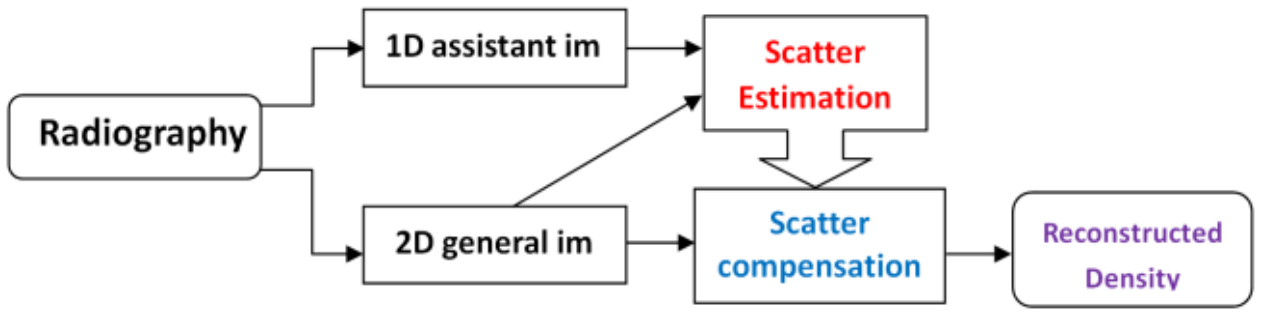

Fig.6. "Constant" scatter correction within reconstruction procedure

For the cases that scatter distribution is not uniform, a scatter correction algorithm is developed which structure is shown in Fig.7. Part of structure in gray background is same to scatter estimation procedure given in Fig.6. With this scatter estimation procedure, a "real" scatter value of certain position of visible image can be measured. Simulation of scatter distribution on the image is provided by Monte Carlo. Then a scale constant C can be obtained, with which simulation distribution can be transferred to an artificial measured one.

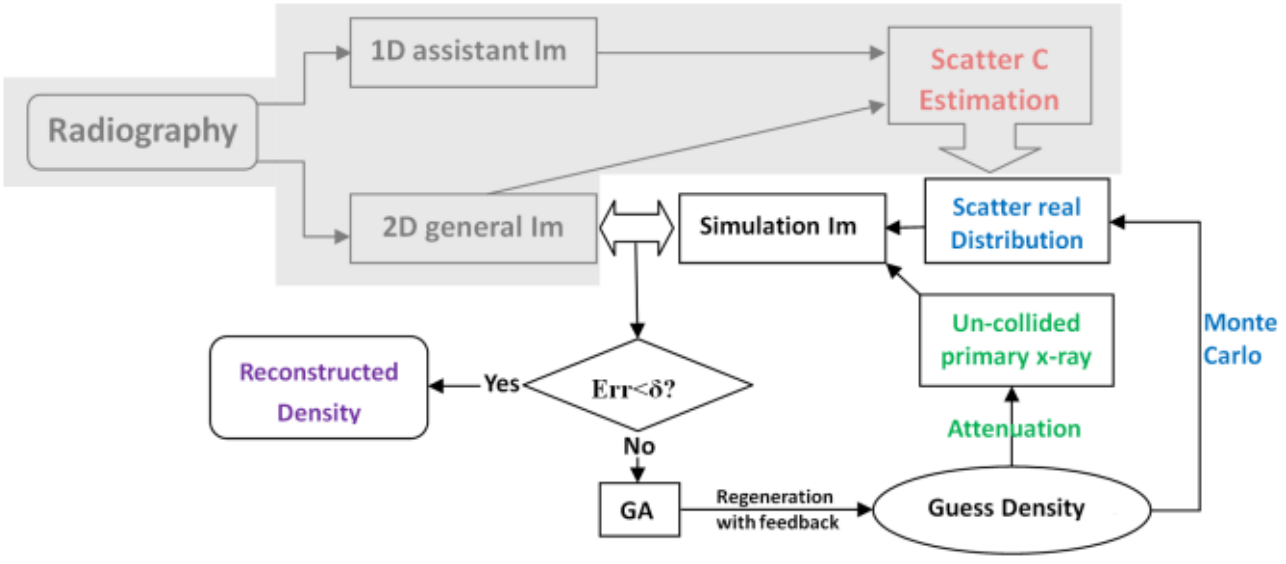

Fig.7. General scatter correction within reconstruction procedure

The rest part with on background shows an iterative improvement reconstruction procedure. With a initial guess of object density, radiography simulation image with scatter and primary x-ray is obtained. Then a direct comparison between simulation and captured visible image is made using scale constant C. Difference given by the comparison is feedback to an intelligent optimizer to improve the guess density. Iteration stopped, once the error is smaller than a threshold value δ. Then the final guess density is suggested as reconstructed result.

Here, Genetic algorithm GA[14] is employed as intelligent optimizer to searching the best density which projection matches experimental visible image. Although GA is not fast, it is the most robust algorithm for its global convergence property. In addition to this, no regularization and related parameters are introduced which impose artificial smoothness and edge preserving on reconstructed image.

Since, Monte Carlo simulations in iteration cost a big computational expense. Acceleration methods of simulation are studied. Reference[15] suggests that down-sampling the projection data and image is a good way to reduce the computational expense. It is known that scatter distribution is more smooth on FOV than that of primary x-ray. Primary and scattered x-ray are simulated by two different methods. For an object at cylinder or sphere coordinate, attenuation of primary x-ray can be calculated accurately according to interactions of its characteristic ray and object layers. Detail study can be found in reference [16].

Monte Carlo code JMCT[17] serves as simulation tool for scattered x-ray. CPU consumption is proportion to the number of pixels at which the scattered x-ray are calculated. Enlightened by article[18], an adaptive down-sampling method is used to reduce CPU time. Procedure of this adaptive down-sampling method is as follows. According to the distribution of scatter in reconstruction iterations, not all but only several pixels on the image are selected. For example, in Fig.8, the scatter is flatly distributed near the central of FOV, while it changes a lot at the region away from center. The number of tally points about scatter on the image is down-sampling to six. Two points near and four points away from the FOV center are selected according to the rough distribution shape in iterations. Then simulated results of these six pixels are fitted and up-sampled by Possion-kriging[19] interpolation to the initial resolution scale, and scatter distribution is then obtained. As shown in Fig.8, distributions given by this adaptive down-sampling method and direct Monte Carlo simulation pixel by pixel are in good agreement.

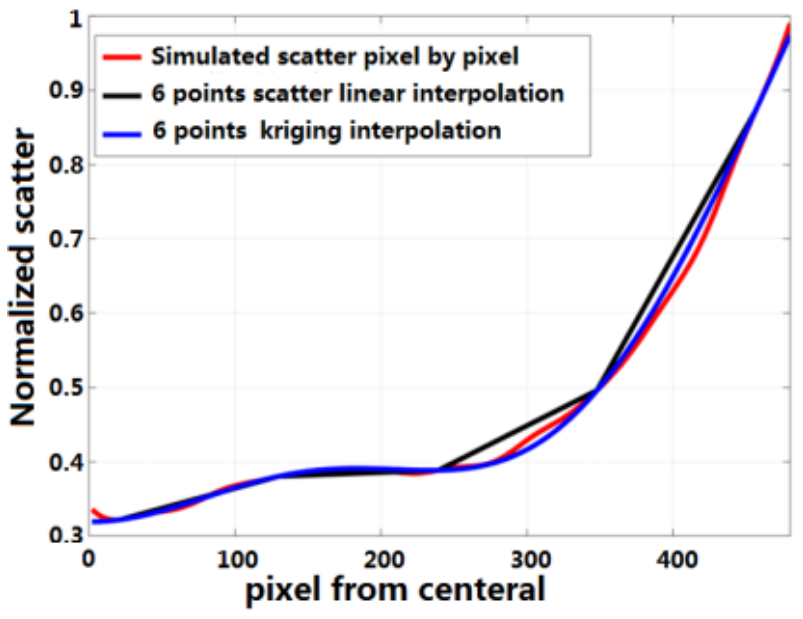

Fig.8 Simulated scatter and six points scatter interpolation

## IV. Results and Discussion

With the proposed iterative reconstruction procedure presented in Fig6, the simulated FTO object projection as shown in fig.2.a. is reconstructed for 20 times. All results named Ser-1 to Ser-20 are given in Fig.9. It is clear that average of 20 densities agrees with the designed value. The differences between 20 densities are at the layers with 1~2 cm radius. It is because that, the less intensity of x-ray, the more sensitively that reconstructed density is effected by scatter. As shown in Fig.2 the intensity of primary x-ray reaches its minimum corresponding to projection of these layers, because, the characteristic ray of primary x-ray which is tangential these layers has the longest optical length.

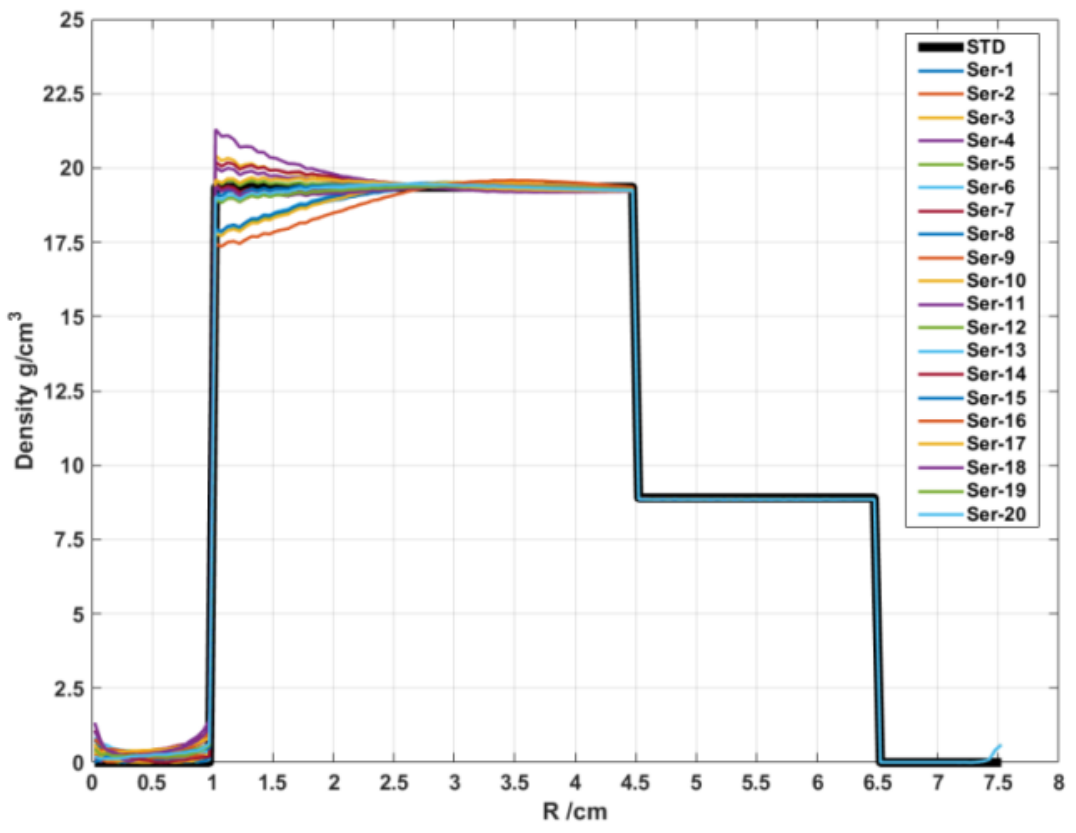

Fig.9. Reconstructed densities and standard one.

Another in-homogeneous object is used to show the performance as well. Its density distribution named Origin(STD) is shown in Fig.10. The maximum scatter-to-primary ratio is over 30, which means at this region of visible image the proportion of useful primary information is less than 3.5%. Here, reconstructions by GA( shown in Fig.7) are carried out 100 times. The obtained maximum and minimum densities are given as GA-up and GA-down in Fig.10. The average of reconstructed 50 densities is given with name GA as well. In addition, with reconstruction procedure shown in Fig.6, density is produced with name SE in Fig.10.

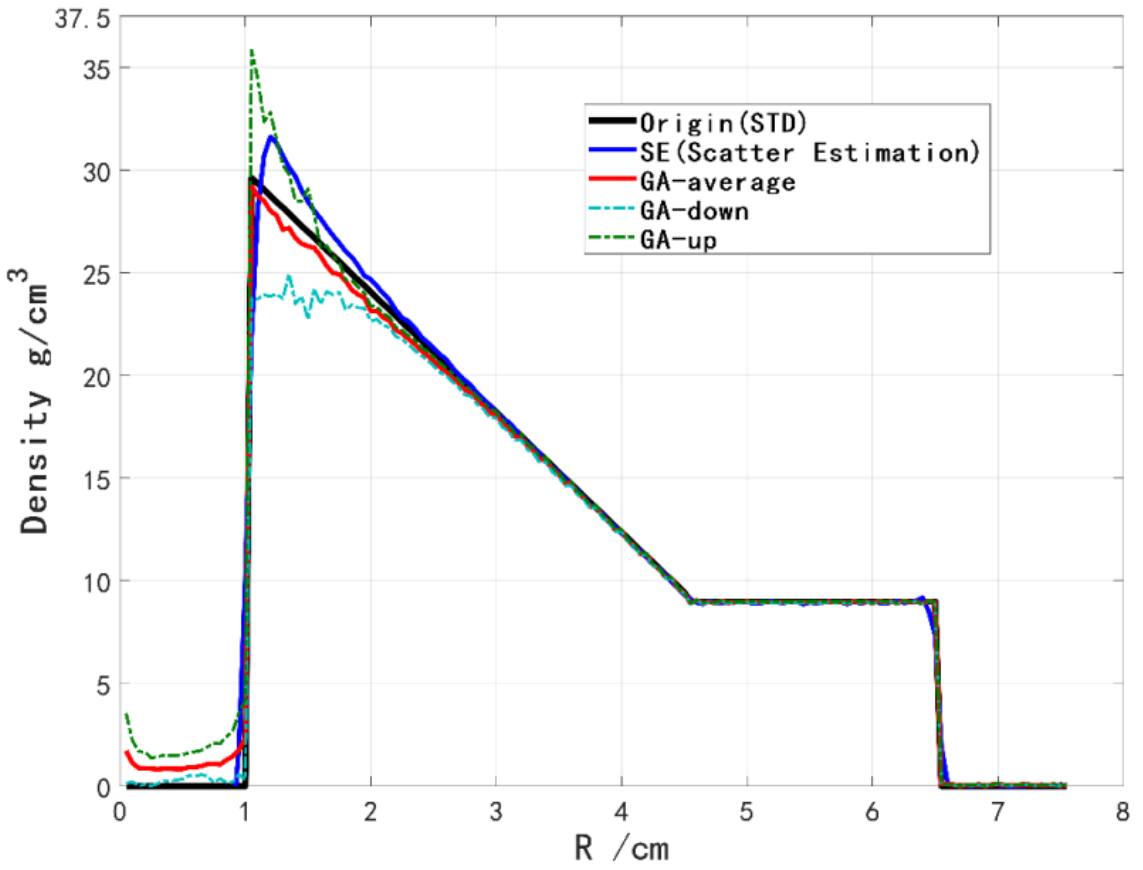

Fig.10. Density of in-homogeneous object(Origin) and reconstructed densities with both Scatter estimation(shown in Fig.6) and GA algorithm (shown in Fig.7).

A comparison between GA, SE and Origin shows that, density given by GA best matches the Origin one with pixel error less than 2%, even though the scatter-to-primary ratio over 30. SE method has error less than 10%. The results of GA-up and GA-down offer a margin to describe uncertainty of the average density GA.

## V. Conclusion

In this paper, scatter correction method for Multi-MeV flash radiography is investigated. An online scatter estimation system is firstly proposed. It can be easily prepared by setting an assistant 1-D imager downstream a general imaging system. Monte Carlo simulations show that this method produces scatter estimation in central of FOV is less than 2% for FTO-U35 object. For the cases that scatter is distributed flatly, the estimated average scatter should suffice density reconstruction. As to universal applications that scatter distribution is not uniform, an reconstruction procedure is designed which combines intelligence algorithm GA for iterative improvement and Monte Carlo for scatter distribution simulation. The proposed online scatter estimation gives the scale constant C linking simulated distribution and experimental one. Radiography simulations about two objects show our reconstruction procedure has a promising performance on scatter correction for density reconstruction.